\documentclass[fleqn]{llncs}
\usepackage{isij}

\title{Instruction Sequences with Indirect Jumps%
       \thanks{This research was partly carried out in the framework of
               the  Jacquard-project Symbiosis, which is funded by the
               Netherlands Organisation for Scientific Research (NWO).}}
\author{J.A. Bergstra\inst{1}\fnmsep\inst{2}
        \and
        C.A. Middelburg\inst{1}
       }
\institute{Programming Research Group,
           University of Amsterdam, \\
           P.O.~Box~41882, 1009~DB~Amsterdam, the Netherlands \\
           \and
           Department of Philosophy,
           Utrecht University, \\
           P.O.~Box~80126, 3508~TC~Utrecht, the Netherlands \\
           \email{J.A.Bergstra@uva.nl,C.A.Middelburg@uva.nl}
          }

\begin{document}
\maketitle

\begin{abstract}
We study sequential programs that are instruction sequences with direct
and indirect jump instructions.
The intuition is that indirect jump instructions are jump instructions
where the position of the instruction to jump to is the content of some
memory cell.
We consider several kinds of indirect jump instructions.
For each kind, we define the meaning of programs with indirect jump
instructions of that kind by means of a translation into programs
without indirect jump instructions.
For each kind, the intended behaviour of a program with indirect jump
instructions of that kind under execution is the behaviour of the
translated program under execution on interaction with some memory
device.
\\[1.5ex]
\\[1.5ex]
{\sl 1998 ACM Computing Classification:}
D.3.1, D.3.3, F.1.1, F.3.2, F.3.3.
\end{abstract}

\section{Introduction}
\label{sect-intro}

We take the view that sequential programs are in essence sequences of
instructions.
Although finite state programs with direct and indirect jump
instructions are as expressive as finite state programs with direct jump
instructions only, indirect jump instructions are widely used.
For example, return instructions, in common use to implement recursive
method calls in programming language such as Java~\cite{GJSB00a} and
C\#~\cite{HWG03a}, are indirect jump instructions.
Therefore, we consider a theoretical understanding of both direct jump
instructions and indirect jump instructions highly relevant to
programming.
In~\cite{BL02a}, sequential programs that are instruction sequences with
direct jump instructions are studied.
In this paper, we study sequential programs that are instruction
sequences with both direct jump instructions and indirect jump
instructions.

We believe that interaction with components of an execution environment,
in particular memory devices, is inherent in the behaviour of programs
under execution.
Intuitively, indirect jump instructions are jump instructions where the
position of the instruction to jump to is the content of some memory
cell.
In this paper, we consider several kinds of indirect jump instructions,
including return instructions.
For each kind, we define the meaning of programs with indirect jump
instructions of that kind by means of a translation into programs
without indirect jump instructions.
For each kind, the intended behaviour of a program with indirect jump
instructions of that kind under execution is the behaviour of the
translated program under execution on interaction with some memory
device.
We also describe the memory devices concerned, to wit register files and
stacks.

The approach to define the meaning of programs mentioned above is
introduced under the name projection semantics in~\cite{BL02a}.
Projection semantics explains the meaning of programs in terms of known
programs instead of more or less sophisticated mathematical objects that
represent behaviours.
The main advantage of projection semantics is that it does not require a
lot of mathematical background.
In the present case, another advantage of projection semantics is that
it follows immediately that indirect jump instructions of the kinds
considered can be eliminated from programs in the presence of an
appropriate memory device.
We will study sequential programs that are instruction sequences with
direct and indirect jump instructions in the setting in which projection
semantics has been developed so far: the setting of program algebra and
basic thread algebra.%
\footnote
{In~\cite{BL02a}, basic thread algebra is introduced under the
 name basic polarized process algebra.
 Prompted by the development of thread algebra~\cite{BM04c}, which is a
 design on top of it, basic polarized process algebra has been renamed
 to basic thread algebra.
}

Program algebra is an algebra of deterministic sequential programs based
on the idea that such programs are in essence sequences of instructions.
Basic thread algebra is a form of process algebra which is tailored to
the description of the behaviour of deterministic sequential programs
under execution.
A hierarchy of program notations rooted in program algebra is introduced
in~\cite{BL02a}.
In this paper, we embroider on two program notations that belong to this
hierarchy.
The program notations in question, called \PGLC\ and \PGLD, are close to
existing assembly languages.
The main difference between them is that \PGLC\ has relative jump
instructions and \PGLD\ has absolute jump instructions.

A thread proceeds by doing steps in a sequential fashion.
A thread may do certain steps only for the sake of having itself
affected by some service.
In~\cite{BP02a}, the use mechanism is introduced to allow for such
interaction between threads and services.
The interaction between behaviours of programs under execution and some
memory device referred to above is an interaction of this kind.
In this paper, we will use a slightly adapted form of the use mechanism,
called thread-service composition, to have behaviours of programs under
execution affected by services.

This paper is organized as follows.
First, we review basic thread algebra, program algebra, and the program
notations \PGLC\ and \PGLD\
(Sections~\ref{sect-BTA}, \ref{sect-PGA}, and~\ref{sect-PGLC-PGLD}).
Next, we extend basic thread algebra with thread-service composition and
introduce a state-based approach to describe services
(Sections~\ref{sect-TAtsc} and~\ref{sect-service-descr}).
Following this, we give a state-based description of register file
services and introduce variants of the program notations \PGLC\ and
\PGLD\ with indirect jump instructions
(Sections~\ref{sect-reg-file}, \ref{sect-PGLDij}, and~\ref{sect-PGLCij}).
We also introduce a variant of one of those program notations with
double indirect jump instructions (Section~\ref{sect-PGLDdij}).
After that, we give a state-based description of stack services and
introduce a variant of the program notation \PGLD\ with returning jump
instructions and return instructions
(Sections~\ref{sect-stack} and~\ref{sect-PGLDrj}).
Finally, we make some concluding remarks (Section~\ref{sect-concl}).

\section{Basic Thread Algebra}
\label{sect-BTA}

In this section, we review \BTA\ (Basic Thread Algebra), a form of
process algebra which is tailored to the description of the behaviour of
deterministic sequential programs under execution.
The behaviours concerned are called \emph{threads}.

In \BTA, it is assumed that there is a fixed but arbitrary finite set of
\emph{basic actions} $\BAct$.
The intuition is that each basic action performed by a thread is taken
as a command to be processed by a service provided by the execution
environment of the thread.
The processing of a command may involve a change of state of the service
concerned.
At completion of the processing of the command, the service produces a
reply value.
This reply is either $\True$ or $\False$ and is returned to the thread
concerned.

Although \BTA\ is one-sorted, we make this sort explicit.
The reason for this is that we will extend \BTA\ with an additional sort
in Section~\ref{sect-TAtsc}.

The algebraic theory \BTA\ has one sort: the sort $\Thr$ of
\emph{threads}.
To build terms of sort $\Thr$, \BTA\ has the following constants and
operators:
\begin{iteml}
\item
the \emph{deadlock} constant $\const{\DeadEnd}{\Thr}$;
\item
the \emph{termination} constant $\const{\Stop}{\Thr}$;
\item
for each $a \in \BAct$, the binary \emph{postconditional composition}
operator\linebreak $\funct{\pcc{\ph}{a}{\ph}}{\Thr \x \Thr}{\Thr}$.
\end{iteml}
Terms of sort $\Thr$ are built as usual (see e.g.~\cite{ST99a,Wir90a}).
Throughout the paper, we assume that there are infinitely many variables
of sort $\Thr$, including $x,y,z$.

We use infix notation for postconditional composition.
We introduce \emph{action prefixing} as an abbreviation: $a \bapf p$,
where $p$ is a term of sort $\Thr$, abbreviates $\pcc{p}{a}{p}$.

Let $p$ and $q$ be closed terms of sort $\Thr$ and $a \in \BAct$.
Then $\pcc{p}{a}{q}$ will perform action $a$, and after that proceed as
$p$ if the processing of $a$ leads to the reply $\True$ (called a
positive reply), and proceed as $q$ if the processing of $a$ leads to
the reply $\False$ (called a negative reply).

Each closed \BTA\ term of sort $\Thr$ denotes a finite thread, i.e.\ a
thread of which the length of the sequences of actions that it can
perform is bounded.
Guarded recursive specifications give rise to infinite threads.

A \emph{guarded recursive specification} over \BTA\ is a set of
recursion equations $E = \set{X = t_X \where X \in V}$, where $V$ is a
set of variables of sort $\Thr$ and each $t_X$ is a term of the form
$\DeadEnd$, $\Stop$ or $\pcc{t}{a}{t'}$ with $t$ and $t'$ \BTA\ terms of
sort $\Thr$ that contain only variables from $V$.
We write $\vars(E)$ for the set of all variables that occur on the
left-hand side of an equation in $E$.
We are only interested in models of \BTA\ in which guarded recursive
specifications have unique solutions, such as the projective limit model
of \BTA\ presented in~\cite{BB03a}.
A thread that is the solution of a finite guarded recursive
specification over \BTA\ is called a \emph{finite-state} thread.

We extend \BTA\ with guarded recursion by adding constants for solutions
of guarded recursive specifications and axioms concerning these
additional constants.
For each guarded recursive specification $E$ and each $X \in \vars(E)$,
we add a constant of sort $\Thr$ standing for the unique solution of $E$
for $X$ to the constants of \BTA.
The constant standing for the unique solution of $E$ for $X$ is denoted
by $\rec{X}{E}$.
Moreover, we add the axioms for guarded recursion given in
Table~\ref{axioms-rec} to \BTA,%
\begin{table}[!t]
\caption{Axioms for guarded recursion}
\label{axioms-rec}
\begin{eqntbl}
\begin{saxcol}
\rec{X}{E} = \rec{t_X}{E} & \mif X \!=\! t_X \in E       & \axiom{RDP}
\\
E \Implies X = \rec{X}{E} & \mif X \in \vars(E)          & \axiom{RSP}
\end{saxcol}
\end{eqntbl}
\end{table}
where we write $\rec{t_X}{E}$ for $t_X$ with, for all $Y \in \vars(E)$,
all occurrences of $Y$ in $t_X$ replaced by $\rec{Y}{E}$.
In this table, $X$, $t_X$ and $E$ stand for an arbitrary variable of
sort $\Thr$, an arbitrary \BTA\ term of sort $\Thr$ and an arbitrary
guarded recursive specification over \BTA, respectively.
Side conditions are added to restrict the variables, terms and guarded
recursive specifications for which $X$, $t_X$ and $E$ stand.
The equations $\rec{X}{E} = \rec{t_X}{E}$ for a fixed $E$ express that
the constants $\rec{X}{E}$ make up a solution of $E$.
The conditional equations $E \Implies X = \rec{X}{E}$ express that this
solution is the only one.

We will write \BTA+\REC\ for \BTA\ extended with the constants for
solutions of guarded recursive specifications and axioms RDP and RSP.

In~\cite{BM05c}, we show that the threads considered in \BTA+\REC\ can
be viewed as processes that are definable over ACP~\cite{Fok00}.

\section{Program Algebra}
\label{sect-PGA}

In this section, we review \PGA\ (ProGram Algebra), an algebra of
sequential programs based on the idea that sequential programs are in
essence sequences of instructions.
\PGA\ provides a program notation for finite-state threads.

In \PGA, it is assumed that there is a fixed but arbitrary finite set
$\BInstr$ of \emph{basic instructions}.
\PGA\ has the following \emph{primitive instructions}:
\begin{iteml}
\item
for each $a \in \BInstr$, a \emph{plain basic instruction} $a$;
\item
for each $a \in \BInstr$, a \emph{positive test instruction} $\ptst{a}$;
\item
for each $a \in \BInstr$, a \emph{negative test instruction} $\ntst{a}$;
\item
for each $l \in \Nat$, a \emph{forward jump instruction} $\fjmp{l}$;
\item
a \emph{termination instruction} $\halt$.
\end{iteml}
We write $\PInstr$ for the set of all primitive instructions.

The intuition is that the execution of a basic instruction $a$ may
modify a state and produces $\True$ or $\False$ at its completion.
In the case of a positive test instruction $\ptst{a}$, basic instruction
$a$ is executed and execution proceeds with the next primitive
instruction if $\True$ is produced and otherwise the next primitive
instruction is skipped and execution proceeds with the primitive
instruction following the skipped one.
In the case where $\True$ is produced and there is not at least one
subsequent primitive instruction and in the case where $\False$ is
produced and there are not at least two subsequent primitive
instructions, deadlock occurs.
In the case of a negative test instruction $\ntst{a}$, the role of the
value produced is reversed.
In the case of a plain basic instruction $a$, the value produced is
disregarded: execution always proceeds as if $\True$ is produced.
The effect of a forward jump instruction $\fjmp{l}$ is that execution
proceeds with the $l$-th next instruction of the program concerned.
If $l$ equals $0$ or the $l$-th next instruction does not exist, then
$\fjmp{l}$ results in deadlock.
The effect of the termination instruction $\halt$ is that execution
terminates.

\PGA\ has the following constants and operators:
\begin{iteml}
\item
for each $u \in \PInstr$, an \emph{instruction} constant $u$\,;
\item
the binary \emph{concatenation} operator $\ph \conc \ph$\,;
\item
the unary \emph{repetition} operator $\ph\rep$\,.
\end{iteml}
Terms are built as usual.
Throughout the paper, we assume that there are infinitely many
variables, including $x,y,z$.

We use infix notation for concatenation and postfix notation for
repetition.

Closed \PGA\ terms are considered to denote programs.
The intuition is that a program is in essence a non-empty, finite or
infinite sequence of primitive instructions.
These sequences are called \emph{single pass instruction sequences}
because \PGA\ has been designed to enable single pass execution of
instruction sequences: each instruction can be dropped after it has been
executed.
Programs are considered to be equal if they represent the same single
pass instruction sequence.
The axioms for instruction sequence equivalence are given in
Table~\ref{axioms-PGA}.%
\begin{table}[!t]
\caption{Axioms of \PGA}
\label{axioms-PGA}
\begin{eqntbl}
\begin{axcol}
(x \conc y) \conc z = x \conc (y \conc z)              & \axiom{PGA1} \\
(x^n)\rep = x\rep                                      & \axiom{PGA2} \\
x\rep \conc y = x\rep                                  & \axiom{PGA3} \\
(x \conc y)\rep = x \conc (y \conc x)\rep              & \axiom{PGA4}
\end{axcol}
\end{eqntbl}
\end{table}
In this table, $n$ stands for an arbitrary natural number greater than
$0$.
For each $n > 0$, the term $x^n$ is defined by induction on $n$ as
follows: $x^1 = x$ and $x^{n+1} = x \conc x^n$.
The \emph{unfolding} equation $x\rep = x \conc x\rep$ is
derivable.
Each closed \PGA\ term is derivably equal to a term in
\emph{canonical form}, i.e.\ a term of the form $P$ or $P \conc Q\rep$,
where $P$ and $Q$ are closed \PGA\ terms that do not contain the
repetition operator.

Each closed \PGA\ term is considered to denote a program of which the
behaviour is a finite-state thread, taking the set $\BInstr$ of basic
instructions for the set $\BAct$ of actions.
The \emph{thread extraction} operator $\extr{\ph}$ assigns a thread to
each program.
The thread extraction operator is defined by the equations given in
Table~\ref{axioms-thread-extr} (for $a \in \BInstr$, $l \in \Nat$ and
$u \in \PInstr$)%
\begin{table}[!t]
\caption{Defining equations for thread extraction operator}
\label{axioms-thread-extr}
\begin{eqntbl}
\begin{eqncol}
\extr{a} = a \bapf \DeadEnd \\
\extr{a \conc x} = a \bapf \extr{x} \\
\extr{\ptst{a}} = a \bapf \DeadEnd \\
\extr{\ptst{a} \conc x} =
\pcc{\extr{x}}{a}{\extr{\fjmp{2} \conc x}} \\
\extr{\ntst{a}} = a \bapf \DeadEnd \\
\extr{\ntst{a} \conc x} =
\pcc{\extr{\fjmp{2} \conc x}}{a}{\extr{x}}
\end{eqncol}
\qquad
\begin{eqncol}
\extr{\fjmp{l}} = \DeadEnd \\
\extr{\fjmp{0} \conc x} = \DeadEnd \\
\extr{\fjmp{1} \conc x} = \extr{x} \\
\extr{\fjmp{l+2} \conc u} = \DeadEnd \\
\extr{\fjmp{l+2} \conc u \conc x} = \extr{\fjmp{l+1} \conc x} \\
\extr{\halt} = \Stop \\
\extr{\halt \conc x} = \Stop
\end{eqncol}
\end{eqntbl}
\end{table}
and the rule given in Table~\ref{rule-thread-extr}.%
\begin{table}[!t]
\caption{Rule for cyclic jump chains}
\label{rule-thread-extr}
\begin{eqntbl}
\begin{eqncol}
x \scongr \fjmp{0} \conc y \Implies \extr{x} = \DeadEnd
\end{eqncol}
\end{eqntbl}
\end{table}
This rule is expressed in terms of the \emph{structural congruence}
predicate $\ph \scongr \ph$, which is defined by the formulas given in
Table~\ref{axioms-scongr} (for $n,m,l \in \Nat$ and
$u_1,\ldots,u_n,v_1,\ldots,v_{m+1} \in \PInstr$).%
\begin{table}[!t]
\caption{Defining formulas for structural congruence predicate}
\label{axioms-scongr}
\begin{eqntbl}
\begin{eqncol}
\fjmp{n+1} \conc u_1 \conc \ldots \conc u_n \conc \fjmp{0}
\scongr
\fjmp{0} \conc u_1 \conc \ldots \conc u_n \conc \fjmp{0}
\\
\fjmp{n+1} \conc u_1 \conc \ldots \conc u_n \conc \fjmp{m}
\scongr
\fjmp{m+n+1} \conc u_1 \conc \ldots \conc u_n \conc \fjmp{m}
\\
(\fjmp{n+l+1} \conc u_1 \conc \ldots \conc u_n)\rep \scongr
(\fjmp{l} \conc u_1 \conc \ldots \conc u_n)\rep
\\
\fjmp{m+n+l+2} \conc u_1 \conc \ldots \conc u_n \conc
(v_1 \conc \ldots \conc v_{m+1})\rep \scongr {} \\ \hfill
\fjmp{n+l+1} \conc u_1 \conc \ldots \conc u_n \conc
(v_1 \conc \ldots \conc v_{m+1})\rep
\\
x \scongr x
\\
x_1 \scongr y_1 \And x_2 \scongr y_2 \Implies
x_1 \conc x_2 \scongr y_1 \conc y_2 \And
{x_1}\rep \scongr {y_1}\rep
\end{eqncol}
\end{eqntbl}
\end{table}

The equations given in Table~\ref{axioms-thread-extr} do not cover the
case where there is a cyclic chain of forward jumps.
Programs are structural congruent if they are the same after removing
all chains of forward jumps in favour of single jumps.
Because a cyclic chain of forward jumps corresponds to $\fjmp{0}$,
the rule from Table~\ref{rule-thread-extr} can be read as follows:
if $x$ starts with a cyclic chain of forward jumps, then $\extr{x}$
equals $\DeadEnd$.
It is easy to see that the thread extraction operator assigns the same
thread to structurally congruent programs.
Therefore, the rule from Table~\ref{rule-thread-extr} can be replaced by
the following generalization:
$x \scongr y  \Implies \extr{x} = \extr{y}$.

Let $E$ be a finite guarded recursive specification over \BTA, and let
$P_X$ be a closed \PGA\ term for each $X \in \vars(E)$.
Let $E'$ be the set of equations that results from replacing in $E$ all
occurrences of $X$ by $\extr{P_X}$ for each $X \in \vars(E)$.
If $E'$ can be obtained by applications of axioms PGA1--PGA4, the
defining equations for the thread extraction operator and the rule for
cyclic jump chains, then $\extr{P_X}$ is the solution of $E$ for $X$.
Such a finite guarded recursive specification can always be found.
%
%
Thus, the behaviour of each closed \PGA\ term, is a thread that is
definable by a finite guarded recursive specification over \BTA.
Moreover, each finite guarded recursive specification over \BTA\ can be
translated to a closed \PGA\ term of which the behaviour is the solution
of the finite guarded recursive specification concerned.

Closed \PGA\ terms are loosely called \PGA\ \emph{programs}.
\PGA\ programs in which the repetition operator do not occur
are called \emph{finite} \PGA\ programs.

\section{The Program Notations \PGLC\ and \PGLD}
\label{sect-PGLC-PGLD}

In this section, we review two program notations which are rooted in
\PGA.
These program notations, called \PGLC\ and PGLD, belong to a hierarchy
of program notations introduced in~\cite{BL02a}.

Both \PGLC\ and \PGLD\ are close to existing assembly languages.
The main difference between them is that \PGLC\ has relative jump
instructions and \PGLD\ has absolute jump instructions.
\PGLC\ and \PGLD\ have no explicit termination instruction.

In \PGLC\ and \PGLD, like in \PGA, it is assumed that there is a fixed
but arbitrary set of \emph{basic instructions} $\BInstr$.
Again, the intuition is that the execution of a basic instruction $a$
may modify a state and produces $\True$ or $\False$ at its completion.

\PGLC\ has the following primitive instructions:
\begin{iteml}
\item
for each $a \in \BInstr$, a \emph{plain basic instruction} $a$;
\item
for each $a \in \BInstr$, a \emph{positive test instruction} $\ptst{a}$;
\item
for each $a \in \BInstr$, a \emph{negative test instruction} $\ntst{a}$;
\item
for each $l \in \Nat$, a \emph{direct forward jump instruction}
$\fjmp{l}$;
\item
for each $l \in \Nat$, a \emph{direct backward jump instruction}
$\bjmp{l}$.
\end{iteml}
\PGLC\ programs have the form $u_1 \conc \ldots \conc u_k$, where
$u_1,\ldots,u_k$ are primitive instructions of \PGLC.

The plain basic instructions, the positive test instructions, and the
negative test instructions are as in \PGA, except that termination
instead of deadlock occurs in the case where there are insufficient
subsequent primitive instructions.
The effect of a direct forward jump instruction $\fjmp{l}$ is that
execution proceeds with the $l$-th next instruction of the program
concerned.
If $l$ equals $0$, then deadlock occurs.
If the $l$-th next instruction does not exist, then termination occurs.
The effect of a direct backward jump instruction $\bjmp{l}$ is that
execution proceeds with the $l$-th previous instruction of the program
concerned.
If $l$ equals $0$, then deadlock occurs.
If the $l$-th previous instruction does not exist, then termination
occurs.

We define the meaning of \PGLC\ programs by means of a function
$\pglcpga$ from the set of all \PGLC\ programs to the set of all \PGA\
programs.
This function is defined by
\begin{ldispl}
\pglcpga(u_1 \conc \ldots \conc u_k) =
(\psi_1(u_1) \conc \ldots \conc \psi_k(u_k) \conc
 \halt \conc \halt)\rep\;,
\end{ldispl}%
where the auxiliary functions $\psi_j$ from the set of all primitive
instructions of \PGLC\ to the set of all primitive instructions of \PGA\
are defined as follows ($1 \leq j \leq k$):
\pagebreak[2]
\begin{ldispl}
\begin{aceqns}
\psi_j(\fjmp{l}) & = & \fjmp{l}     & \mif j + l \leq k\;, \\
\psi_j(\fjmp{l}) & = & \halt        & \mif j + l   >  k\;, \\
\psi_j(\bjmp{l}) & = & \fjmp{k+2-l} & \mif l   <  j\;, \\
\psi_j(\bjmp{l}) & = & \halt        & \mif l \geq j\;, \\
\psi_j(u)        & = & u
                    & \mif u\; \mathrm{is\;not\;a\;jump\;instruction}\;.
\end{aceqns}
\end{ldispl}%
The idea is that each backward jump can be replaced by a forward jump if
the entire program is repeated.
To enforce termination of the program after execution of its last
instruction if the last instruction is a plain basic instruction, a
positive test instruction or a negative test instruction,
$\halt \conc \halt$ is appended to
$\psi_1(u_1) \conc \ldots \conc \psi_k(u_k)$.

Let $P$ be a \PGLC\ program.
Then $\pglcpga(P)$ represents the meaning of $P$ as a \PGA\ program.
The intended behaviour of $P$ is the behaviour of $\pglcpga(P)$.
That is, the \emph{behaviour} of $P$, written $\extr{P}_\sPGLC$, is
$\extr{\pglcpga(P)}$.

\PGLD\ has the following primitive instructions:
\begin{itemize}
\item
for each $a \in \BInstr$, a \emph{plain basic instruction} $a$;
\item
for each $a \in \BInstr$, a \emph{positive test instruction} $\ptst{a}$;
\item
for each $a \in \BInstr$, a \emph{negative test instruction} $\ntst{a}$;
\item
for each $l \in \Nat$, a \emph{direct absolute jump instruction}
$\ajmp{l}$.
\end{itemize}
\PGLD\ programs have the form $u_1;\ldots;u_k$, where $u_1,\ldots,u_k$
are primitive instructions of \PGLD.

The plain basic instructions, the positive test instructions, and the
negative test instructions are as in \PGLC.
The effect of a direct absolute jump instruction $\ajmp{l}$ is that
execution proceeds with the $l$-th instruction of the program concerned.
If $\ajmp{l}$ is itself the $l$-th instruction, then deadlock occurs.
If $l$ equals $0$ or $l$ is greater than the length of the program, then
termination occurs.

We define the meaning of \PGLD\ programs by means of a function
$\pgldpglc$ from the set of all \PGLD\ programs to the set of all \PGLC\
programs.
This function is defined by
\begin{ldispl}
\pgldpglc(u_1 \conc \ldots \conc u_k) =
\psi_1(u_1) \conc \ldots \conc \psi_k(u_k)\;,
\end{ldispl}%
where the auxiliary functions $\psi_j$ from the set of all primitive
instructions of \PGLD\ to the set of all primitive instructions of \PGLC\
are defined as follows ($1 \leq j \leq k$):
\begin{ldispl}
\begin{aceqns}
\psi_j(\ajmp{l}) & = & \fjmp{l-j} & \mif l \geq j\;, \\
\psi_j(\ajmp{l}) & = & \bjmp{j-l} & \mif l   <  j\;, \\
\psi_j(u)        & = & u
                    & \mif u\; \mathrm{is\;not\;a\;jump\;instruction}\;.
\end{aceqns}
\end{ldispl}%

\sloppy
Let $P$ be a \PGLD\ program.
Then $\pgldpglc(P)$ represents the meaning of $P$ as a \PGLC\ program.
The intended behaviour of $P$ is the behaviour of $\pgldpglc(P)$.
That is, the \emph{behaviour} of $P$, written $\extr{P}_\sPGLD$, is
$\extr{\pgldpglc(P)}_\sPGLC$.

We use the phrase \emph{projection semantics} to refer to the approach
to semantics followed in this section.
The meaning functions $\pglcpga$ and $\pgldpglc$ are called
\emph{projections}.

\PGLC\ and \PGLD\ are very simple program notations.
The hierarchy of program notations introduced in~\cite{BL02a} also
includes a program notation, called \PGLS, that supports structured
programming by offering conditional and loop constructs instead of
(unstructured) jumps.
Each \PGLS\ program can be translated into a semantically equivalent
\PGLD\ program by means of a number of projections.

\section{Interaction of Threads with Services}
\label{sect-TAtsc}

A thread may perform certain actions only for the sake of getting reply
values returned by a service and that way having itself affected by that
service.
In this section, we introduce thread-service composition, which allows
for threads to be affected by services in this way.
We will only use thread-service composition to have program behaviours
affected by a service.
Thread-service composition is a slightly adapted form of the use
mechanism introduced in~\cite{BP02a}.

We consider only deterministic services.
This will do in the case that we address: services that keep private
data for a program.
The services concerned are para-target services by the classification
given in~\cite{BM07a}.

It is assumed that there is a fixed but arbitrary finite set of
\emph{foci} $\Foci$ and a fixed but arbitrary finite set of
\emph{methods} $\Meth$.
Each focus plays the role of a name of a service provided by the
execution environment that can be requested to process a command.
Each method plays the role of a command proper.
For the set $\BAct$ of actions, we take the set
$\set{f.m \where f \in \Foci, m \in \Meth}$.
Performing an action $f.m$ is taken as making a request to the
service named $f$ to process command $m$.

We introduce yet another sort: the sort $\Serv$ of \emph{services}.
However, we will not introduce constants and operators to build terms
of this sort.
$\Serv$ is a parameter of theories with thread-service composition.
$\Serv$ is considered to stand for the set of all services.
It is assumed that each service can be represented by a function
$\funct{H}{\neseqof{\Meth}}{\set{\True,\False,\Blocked}}$ with the
property that
$H(\alpha) = \Blocked \Implies H(\alpha \concat \seq{m}) = \Blocked$ for
all $\alpha \in \neseqof{\Meth}$ and $m \in \Meth$.
This function is called the \emph{reply} function of the service.
Given a reply function $H$ and a method $m \in \Meth$, the
\emph{derived} reply function of $H$ after processing $m$, written
$\derive{m}H$, is defined by
$\derive{m}H(\alpha) = H(\seq{m} \concat \alpha)$.

The connection between a reply function $H$ and the service represented
by it can be understood as follows:
\begin{iteml}
\item
if $H(\seq{m}) = \True$, the request to process command $m$ is accepted
by the service, the reply is positive and the service proceeds as
$\derive{m}H$;
\item
if $H(\seq{m}) = \False$, the request to process command $m$ is accepted
by the service, the reply is negative and the service proceeds as
$\derive{m}H$;
\item
if $H(\seq{m}) = \Blocked$, the request to process command $m$ is
not accepted by the service.
\end{iteml}
Henceforth, we will identify a reply function with the service
represented by it.

For each $f \in \Foci$, we introduce the binary \emph{thread-service
composition} operator $\funct{\use{\ph}{f}{\ph}}{\Thr \x \Serv}{\Thr}$.
Intuitively, $\use{p}{f}{H}$ is the thread that results from processing
all actions performed by thread $p$ that are of the form $f.m$ by
service $H$.
Service $H$ affects thread $p$ by means of the reply values produced at
completion of the processing of the actions performed by $p$.
The actions processed by $H$ are no longer observable.

The axioms for the thread-service composition operator are given in
Table~\ref{axioms-tsc}.%
\begin{table}[!t]
\caption{Axioms for thread-service composition}
\label{axioms-tsc}
\begin{eqntbl}
\begin{saxcol}
\use{\Stop}{f}{H} = \Stop                           & & \axiom{TSC1} \\
\use{\DeadEnd}{f}{H} = \DeadEnd                     & & \axiom{TSC2} \\
\use{(\pcc{x}{g.m}{y})}{f}{H} =
\pcc{(\use{x}{f}{H})}{g.m}{(\use{y}{f}{H})}
                         & \mif f \neq g              & \axiom{TSC3} \\
\use{(\pcc{x}{f.m}{y})}{f}{H} = \use{x}{f}{\derive{m}H}
                         & \mif H(\seq{m}) = \True    & \axiom{TSC4} \\
\use{(\pcc{x}{f.m}{y})}{f}{H} = \use{y}{f}{\derive{m}H}
                         & \mif H(\seq{m}) = \False   & \axiom{TSC5} \\
\use{(\pcc{x}{f.m}{y})}{f}{H} = \DeadEnd
                         & \mif H(\seq{m}) = \Blocked & \axiom{TSC6}
\end{saxcol}
\end{eqntbl}
\end{table}
In this table, $f$ stands for an arbitrary focus from $\Foci$, $m$
stands for an arbitrary method from $\Meth$.
Axiom TSC3 expresses that actions of the form $g.m$, where $f \neq g$,
are not processed.
Axioms TSC4 and TSC5 express that a thread is affected by a service as
described above when an action of the form $f.m$ performed by the thread
is processed by the service.
Axiom TSC6 expresses that deadlock takes place when an action to be
processed is not accepted.

Let $T$ stand for either \BTA\ or \BTA+\REC.
Then we will write $T+\TSC$ for $T$, taking the set
$\set{f.m \where f \in \Foci, m \in \Meth}$ for $\BAct$, extended with
the thread-service composition operators and the axioms from
Table~\ref{axioms-tsc}.

In~\cite{BM05c}, we show that the services considered here can be viewed
as processes that are definable over an extension of \ACP\ with
conditionals introduced in~\cite{BM05a}.

\section{State-Based Description of Services}
\label{sect-service-descr}

In this section, we introduce the state-based approach to describe
families of services that will be used later on.
This approach is similar to the approach to describe state machines
introduced in~\cite{BP02a}.

In this approach, a family of services is described by
\begin{itemize}
\item
a set of states $S$;
\item
an effect function $\funct{\eff}{\Meth \x S}{S}$;
\item
a yield function
$\funct{\yld}{\Meth \x S}{\set{\True,\False,\Blocked}}$;
\end{itemize}
satisfying the following condition:
\begin{ldispl}
\Exists{s \in S}
 {\Forall{m \in \Meth}
   {{} \\ \quad
    (\yld(m,s) = \Blocked \And
     \Forall{s' \in S}
      {(\yld(m,s') = \Blocked \Implies \eff(m,s') = s)})}}\;.
\end{ldispl}%
The set $S$ contains the states in which the services may be; and the
functions $\eff$ and $\yld$ give, for each method $m$ and state $s$, the
state and reply, respectively, that result from processing $m$ in state
$s$.

We define, for each $s \in S$, a cumulative effect function
$\funct{\ceff_s}{\seqof{\Meth}}{S}$ in terms of $s$ and $\eff$ as follows:
\begin{ldispl}
\ceff_s(\emptyseq) = s\;,
\\
\ceff_s(\alpha \concat \seq{m}) = \eff(m,\ceff_s(\alpha))\;.
\end{ldispl}%
We define, for each $s \in S$, a service
$\funct{H_s}{\neseqof{\Meth}}{\set{\True,\False,\Blocked}}$
in terms of $\ceff_s$ and $\yld$ as follows:
\begin{ldispl}
H_s(\alpha \concat \seq{m}) = \yld(m,\ceff_s(\alpha))\;.
\end{ldispl}%
$H_s$ is called the service with \emph{initial state} $s$ described by
$S$, $\eff$ and $\yld$.
We say that $\set{H_s \where s \in S}$ is the \emph{family of services}
described by $S$, $\eff$ and $\yld$.

For each $s \in S$, $H_s$ is a service indeed: the condition imposed on
$S$, $\eff$ and $\yld$ implies that $H_s(\alpha) = \Blocked \Implies
H_s(\alpha \concat \seq{m}) = \Blocked$ for all
$\alpha \in \neseqof{\Meth}$ and $m \in \Meth$.
It is worth mentioning that $H_s(\seq{m}) = \yld(m,s)$ and
$\derive{m} H_s = H_{\eff(m,s)}$.

\section{Register File Services}
\label{sect-reg-file}

In this section, we give a state-based description of the very simple
family of services that constitute a register file of which the
registers can contain natural numbers up to some bound.
This register file will be used in
Sections~\ref{sect-PGLDij}--\ref{sect-PGLDdij} to describe the behaviour
of programs in variants of \PGLC\ and \PGLD\ with indirect jump
instructions.

It is assumed that a fixed but arbitrary number $\maxr$ has been given,
which is considered the number of registers available.
It is also assumed that a fixed but arbitrary number $\maxn$ has been
given, which is considered the greatest natural number that can be
contained in a register.

The register file services accept the following methods:
\begin{itemize}
\item
for each $i \in [0,\maxr]$ and $n \in [0,\maxn]$,
a \emph{register set method} $\setr{:}i{:}n$;
\item
for each $i \in [0,\maxr]$ and $n \in [0,\maxn]$,
a \emph{register test method} $\eqr{:}i{:}n$.
\end{itemize}
We write $\Meth_\rf$ for the set
$\set{\setr{:}i{:}n,\eqr{:}i{:}n \where
      i \in [0,\maxr] \And n \in [0,\maxn]}$.
It is assumed that $\Meth_\rf \subseteq \Meth$.

The methods accepted by register file services can be explained as
follows:
\begin{itemize}
\item
$\setr{:}i{:}n$\,:
the contents of register $i$ becomes $n$ and the reply is $\True$;
\item
$\eqr{:}i{:}n$\,:
if the contents of register $i$ equals $n$, then nothing changes and the
reply is $\True$; otherwise nothing changes and the reply is $\False$.
\end{itemize}

Let $\funct{s}{[1,\maxr]}{[0,\maxn]}$.
Then we write $\RF_s$ for the service with initial state $s$ described
by $S = (\mapof{[1,\maxr]}{[0,\maxn]}) \union \set{\undef}$, where
$\undef \not\in \mapof{[1,\maxr]}{[0,\maxn]}$,\pagebreak[2] and the
functions $\eff$ and $\yld$ defined as follows ($n \in [0,\maxn]$,
$\funct{\rho}{[1,\maxr]}{[0,\maxn]}$):%
\footnote
{We use the following notation for functions:
 $f \owr g$ for the function $h$ with $\dom(h) = \dom(f) \union \dom(g)$
 such that for all $d \in \dom(h)$, $h(d) = f(d)$ if $d \not\in \dom(g)$
 and $h(d) = g(d)$ otherwise; and
 $\maplet{d}{r}$ for the function $f$ with $\dom(f) = \set{d}$ such that
 $f(d) = r$.}%
\begin{ldispl}
\begin{gceqns}
\eff(\setr{:}i{:}n,\rho) = \rho \owr \maplet{i}{n}\;,
\\
\eff(\eqr{:}i{:}n,\rho)  = \rho\;,
\\
\eff(m,\rho)      = \undef       & \mif m \not\in \Meth_\rf\;,
\\
\eff(m,\undef) = \undef\;,
\eqnsep
\yld(\setr{:}i{:}n,\rho) = \True\;,
\\
\yld(\eqr{:}i{:}n,\rho) = \True  & \mif \rho(i) = n\;,
\\
\yld(\eqr{:}i{:}n,\rho) = \False & \mif \rho(i) \neq n\;,
\\
\yld(m,\rho)      = \Blocked     & \mif m \not\in \Meth_\rf\;,
\\
\yld(m,\undef) = \Blocked\;.
\end{gceqns}
\end{ldispl}%
We write $\RF_\mathrm{init}$ for
$\RF_{\maplet{1}{0} \owr \ldots \owr \maplet{I}{0}}$.

\section{\PGLD\ with Indirect Jumps}
\label{sect-PGLDij}

In this section, we introduce a variant of \PGLD\ with indirect jump
instructions.
This variant is called \PGLDij.

In \PGLDij, it is assumed that there is a fixed but arbitrary finite set
of \emph{foci} $\Foci$ with $\rf \in \Foci$ and a fixed but arbitrary
finite set of \emph{methods} $\Meth$.
Moreover, we adopt the assumptions made about register file services in
Section~\ref{sect-reg-file}.
The set $\set{f.m \where f \in \Foci, m \in \Meth}$ is taken as the set
$\BInstr$ of basic instructions.

\PGLDij\ has the following primitive instructions:
\begin{iteml}
\item
for each $a \in \BInstr$, a \emph{plain basic instruction} $a$;
\item
for each $a \in \BInstr$, a \emph{positive test instruction} $\ptst{a}$;
\item
for each $a \in \BInstr$, a \emph{negative test instruction} $\ntst{a}$;
\item
for each $l \in \Nat$, a \emph{direct absolute jump instruction}
$\ajmp{l}$;
\item
for each $i \in [1,\maxr]$, an \emph{indirect absolute jump instruction}
$\iajmp{i}$.
\end{iteml}
\PGLDij\ programs have the form $u_1 \conc \ldots \conc u_k$, where
$u_1,\ldots,u_k$ are primitive instructions of \PGLDij.
\pagebreak[2]

The plain basic instructions, the positive test instructions, the
negative test instructions, and the direct absolute jump instructions
are as in \PGLD.
The effect of an indirect absolute jump instruction $\iajmp{i}$ is that
execution proceeds with the $l$-th instruction of the program concerned,
where $l$ is the content of register $i$.
If $\iajmp{i}$ is itself the $l$-th instruction, then deadlock occurs.
If $l$ equals $0$ or $l$ is greater than the length of the program,
termination occurs.

Recall that the content of register $i$ can be set to $l$ by means of
the basic instruction $\rf.\setr{:}i{:}l$.
Initially, its content is $0$.

Like before, we define the meaning of \PGLDij\ programs by means of a
function $\pgldijpgld$ from the set of all \PGLDij\ programs to the set
of all \PGLD\ programs.
This function is defined by
\begin{ldispl}
\pgldijpgld(u_1 \conc \ldots \conc u_k) = \\ \quad
\psi(u_1) \conc \ldots \conc \psi(u_k) \conc
\ajmp{0} \conc \ajmp{0}  \conc {} \\ \quad
\ptst{\rf.\eqr{:}1{:}1} \conc \ajmp{1} \conc \ldots \conc
\ptst{\rf.\eqr{:}1{:}n} \conc \ajmp{n} \conc \ajmp{0} \conc {} \\
\qquad \vdots  \\ \quad
\ptst{\rf.\eqr{:}\maxr{:}1} \conc \ajmp{1} \conc \ldots \conc
\ptst{\rf.\eqr{:}\maxr{:}n} \conc \ajmp{n} \conc \ajmp{0}\;,
\end{ldispl}%
where $n = \min(k,\maxn)$ and the auxiliary function $\psi$ from the set
of all primitive instructions of \PGLDij\ to the set of all primitive
instructions of \PGLD\ is defined as follows:
\begin{ldispl}
\begin{aceqns}
\psi(\ajmp{l})  & = & \ajmp{l} & \mif l \leq k\;, \\
\psi(\ajmp{l})  & = & \ajmp{0} & \mif l   >  k\;, \\
\psi(\iajmp{i}) & = & \ajmp{l_i}\;, \\
\psi(u) & = & u & \mif u\; \mathrm{is\;not\;a\;jump\;instruction}\;,
\end{aceqns}
\end{ldispl}%
and for each $i \in [1,\maxr]$:
\begin{ldispl}
\begin{aeqns}
l_i & = & k + 3 + (2 \mul \min(k,\maxn) + 1) \mul (i - 1)\;.
\end{aeqns}
\end{ldispl}%
The idea is that each indirect absolute jump can be replaced by a direct
absolute jump to the beginning of the instruction sequence
\begin{ldispl}
\begin{aeqns}
\ptst{\rf.\eqr{:}i{:}1} \conc \ajmp{1} \conc \ldots \conc
\ptst{\rf.\eqr{:}i{:}n} \conc \ajmp{n} \conc \ajmp{0}\;,
\end{aeqns}
\end{ldispl}%
where $i$ is the register concerned and $n = \min(k,\maxn)$.
The execution of this instruction sequence leads to the intended jump
after the content of the register concerned has been found by a linear
search.
To enforce termination of the program after execution of its last
instruction if the last instruction is a plain basic instruction, a
positive test instruction or a negative test instruction,
$\ajmp{0} \conc \ajmp{0}$ is appended to
$\psi(u_1) \conc \ldots \conc \psi(u_k)$.
Because the length of the translated program is greater than $k$, care
is taken that there are no direct absolute jumps to instructions with a
position greater than $k$.
Obviously, the linear search for the content of a register can be
replaced by a binary search.

Let $P$ be a \PGLDij\ program.
Then $\pgldijpgld(P)$ represents the meaning of $P$ as a \PGLD\ program.
The intended behaviour of $P$ is the behaviour of $\pgldijpgld(P)$ on
interaction with a register file.
That is, the \emph{behaviour} of $P$, written $\extr{P}_\sPGLDij$, is
$\use{\extr{\pgldijpgld(P)}_\sPGLD}{\rf}{\RF_\mathrm{init}}$.

More than one instruction is needed in \PGLD\ to obtain the effect of a
single indirect absolute jump instruction.
The projection $\pgldijpgld$ deals with that in such a way that there is
no need for the unit instruction operator introduced in~\cite{Pon02a} or
the distinction between first-level instructions and second-level
instructions introduced in~\cite{BB06a}.

\section{\PGLC\ with Indirect Jumps}
\label{sect-PGLCij}

In this section, we introduce a variant of \PGLC\ with indirect jump
instructions.
This variant is called \PGLCij.

In \PGLCij, the same assumptions are made as in \PGLDij.
Like in \PGLDij, the set $\set{f.m \where f \in \Foci, m \in \Meth}$ is
taken as the set $\BInstr$ of basic instructions.

\PGLDij\ has the following primitive instructions:
\begin{iteml}
\item
for each $a \in \BInstr$, a \emph{plain basic instruction} $a$;
\item
for each $a \in \BInstr$, a \emph{positive test instruction} $\ptst{a}$;
\item
for each $a \in \BInstr$, a \emph{negative test instruction} $\ntst{a}$;
\item
for each $l \in \Nat$, a \emph{direct forward jump instruction}
$\fjmp{l}$;
\item
for each $l \in \Nat$, a \emph{direct backward jump instruction}
$\bjmp{l}$;
\item
for each $i \in [1,\maxr]$, an \emph{indirect forward jump instruction}
$\ifjmp{i}$;
\item
for each $i \in [1,\maxr]$, an \emph{indirect backward jump instruction}
$\ibjmp{i}$.
\end{iteml}
\PGLCij\ programs have the form $u_1 \conc \ldots \conc u_k$, where
$u_1,\ldots,u_k$ are primitive instructions of \PGLCij.

The plain basic instructions, the positive test instructions, the
negative test instructions, the direct forward jump instructions, and
the direct backward jump instructions are as in \PGLC.
The effect of an indirect forward jump instruction $\ifjmp{i}$ is that
execution proceeds with the $l$-th next instruction of the program
concerned, where $l$ is the content of register $i$.
If $l$ equals $0$, then deadlock occurs.
If the $l$-th next instruction does not exist, then termination occurs.
The effect of an indirect backward jump instruction $\ibjmp{i}$ is that
execution proceeds with the $l$-th previous instruction of the program
concerned, where $l$ is the content of register $i$.
If $l$ equals $0$, then deadlock occurs.
If the $l$-th previous instruction does not exist, then termination
occurs.

We define the meaning of \PGLCij\ programs by means of a function
$\pglcijpglc$ from the set of all \PGLCij\ programs to the set of all
\PGLC\ programs.
This function is defined by
\begin{ldispl}
\pglcijpglc(u_1 \conc \ldots \conc u_k) = {} \\ \quad
\psi_1(u_1) \conc \ldots \conc \psi_k(u_k) \conc
\bjmp{k+1} \conc \bjmp{k+2}  \conc {} \\ \quad
\ptst{\rf.\eqr{:}1{:}0} \conc \bjmp{l'_{1,1,0}} \conc
\ldots \conc
\ptst{\rf.\eqr{:}1{:}\maxn} \conc \bjmp{l'_{1,1,\maxn}} \conc {} \\
\qquad \vdots \\ \quad
\ptst{\rf.\eqr{:}1{:}0} \conc \bjmp{l'_{1,k,0}} \conc
\ldots \conc
\ptst{\rf.\eqr{:}1{:}\maxn} \conc \bjmp{l'_{1,k,\maxn}} \conc {}
\eqnsep
\qquad \quad \vdots \eqnsep \quad
\ptst{\rf.\eqr{:}\maxr{:}0} \conc \bjmp{l'_{\maxr,1,0}} \conc
\ldots \conc
\ptst{\rf.\eqr{:}\maxr{:}\maxn} \conc \bjmp{l'_{\maxr,1,\maxn}}
 \conc {} \\
\qquad \vdots \\ \quad
\ptst{\rf.\eqr{:}\maxr{:}0} \conc \bjmp{l'_{\maxr,k,0}} \conc
\ldots \conc
\ptst{\rf.\eqr{:}\maxr{:}\maxn} \conc \bjmp{l'_{\maxr,k,\maxn}}
 \conc {}
\end{ldispl}
\begin{ldispl}
 \quad
\ptst{\rf.\eqr{:}1{:}0} \conc \bjmp{\ul{l}'_{1,1,0}} \conc
\ldots \conc
\ptst{\rf.\eqr{:}1{:}\maxn} \conc \bjmp{\ul{l}'_{1,1,\maxn}}
 \conc {} \\
\qquad \vdots \\ \quad
\ptst{\rf.\eqr{:}1{:}0} \conc \bjmp{\ul{l}'_{1,k,0}} \conc
\ldots \conc
\ptst{\rf.\eqr{:}1{:}\maxn} \conc \bjmp{\ul{l}'_{1,k,\maxn}} \conc {}
\eqnsep
\qquad \quad \vdots \eqnsep \quad
\ptst{\rf.\eqr{:}\maxr{:}0} \conc \bjmp{\ul{l}'_{\maxr,1,0}} \conc
\ldots \conc
\ptst{\rf.\eqr{:}\maxr{:}\maxn} \conc \bjmp{\ul{l}'_{\maxr,1,\maxn}}
 \conc {} \\
\qquad \vdots \\ \quad
\ptst{\rf.\eqr{:}\maxr{:}0} \conc \bjmp{\ul{l}'_{\maxr,k,0}} \conc
\ldots \conc
\ptst{\rf.\eqr{:}\maxr{:}\maxn} \conc
\bjmp{\ul{l}'_{\maxr,k,\maxn}}\;,
\end{ldispl}%
where the auxiliary functions $\psi_j$ from the set of all primitive
instructions of \PGLCij\ to the set of all primitive instructions of
\PGLC\ is defined as follows ($1 \leq j \leq k$):
\begin{ldispl}
\begin{aceqns}
\psi_j(\fjmp{l})  & = & \fjmp{l} & \mif j + l \leq k\;, \\
\psi_j(\fjmp{l})  & = & \bjmp{j} & \mif j + l   >  k\;, \\
\psi_j(\bjmp{l})  & = & \bjmp{l}\;, \\
\psi_j(\ifjmp{i}) & = & \fjmp{l_{i,j}}\;, \\
\psi_j(\ibjmp{i}) & = & \fjmp{\ul{l}_{i,j}}\;, \\
\psi_j(u) & = & u\; & \mif u\; \mathrm{is\;not\;a\;jump\;instruction}\;,
\end{aceqns}
\end{ldispl}%
and for each $i \in [1,\maxr]$, $j \in [1,k]$, and $h \in [0,\maxn]$:
\begin{ldispl}
\begin{aceqns}
l_{i,j} & = &
k+3 + 2 \mul (\maxn+1) \mul (k \mul (i-1) + (j-1))\;, \\
\ul{l}_{i,j} & = &
k+3 + 2 \mul (\maxn+1) \mul (k \mul (\maxr + i-1) + (j-1))\;,
\eqnsep
l'_{i,j,h} & = &
l_{i,j} + 2 \mul h + 1 - (j + h)         & \mif j + h \leq k\;, \\
l'_{i,j,h} & = &
k+3 + 2 \mul (\maxn+1) \mul k \mul \maxr & \mif j + h > k\;,
\eqnsep
\ul{l}'_{i,j,h} & = &
\ul{l}_{i,j} + 2 \mul h + 1 - (j - h)    & \mif j - h \geq 0\;, \\
\ul{l}'_{i,j,h} & = &
k+3 + 4 \mul (\maxn+1) \mul k \mul \maxr & \mif j - h < 0\;.
\end{aceqns}
\end{ldispl}%
Like in the case of indirect absolute jumps, the idea is that each
indirect forward jump and each indirect backward jump can be replaced by
a direct forward jump to the beginning of an instruction sequence whose
execution leads to the intended jump after the content of the register
concerned has been found by a linear search.
However, the direct backward jump instructions occurring in that
instruction sequence now depend upon the position of the indirect jump
concerned in $u_1 \conc \ldots \conc u_k$.
To enforce termination of the program after execution of its last
instruction if the last instruction is a plain basic instruction, a
positive test instruction or a negative test instruction,
$\bjmp{k+1} \conc \bjmp{k+2}$ is appended to
$\psi_1(u_1) \conc \ldots \conc \psi_k(u_k)$.
Because the length of the translated program is greater than $k$, care
is taken that there are no direct forward jumps to instructions with a
position greater than $k$.

Let $P$ be a \PGLCij\ program.
Then $\pglcijpglc(P)$ represents the meaning of $P$ as a \PGLC\ program.
The intended behaviour of $P$ is the behaviour of $\pglcijpglc(P)$ on
interaction with a register file.
That is, the \emph{behaviour} of $P$, written $\extr{P}_\sPGLCij$, is
$\use{\extr{\pglcijpglc(P)}_\sPGLC}{\rf}{\RF_\mathrm{init}}$.

The projection $\pglcijpglc$ yields needlessly long \PGLC\ programs
because it does not take into account the fact that there is at most one
indirect jump instruction at each position in a \PGLCij\ program being
projected.
Taking this fact into account would lead to a projection with a much
more complicated definition.

\section{\PGLD\ with Double Indirect Jumps}
\label{sect-PGLDdij}

In this section, we introduce a variant of \PGLDij\ with double indirect
jump instructions.
This variant is called \PGLDdij.

In \PGLDdij, the same assumptions are made as in \PGLDij.
Like in \PGLDij, the set $\set{f.m \where f \in \Foci, m \in \Meth}$ is
taken as the set $\BInstr$ of basic instructions.

\PGLDdij\ has the following primitive instructions:
\begin{iteml}
\item
for each $a \in \BInstr$, a \emph{plain basic instruction} $a$;
\item
for each $a \in \BInstr$, a \emph{positive test instruction} $\ptst{a}$;
\item
for each $a \in \BInstr$, a \emph{negative test instruction} $\ntst{a}$;
\item
for each $l \in \Nat$, a \emph{direct absolute jump instruction}
$\ajmp{l}$;
\item
for each $i \in [1,\maxr]$, an \emph{indirect absolute jump instruction}
$\iajmp{i}$;
\item
for each $i \in [1,\maxr]$,
a \emph{double indirect absolute jump instruction} $\diajmp{i}$.
\end{iteml}
\PGLDdij\ programs have the form $u_1 \conc \ldots \conc u_k$, where
$u_1,\ldots,u_k$ are primitive instructions of \PGLDdij.

The plain basic instructions, the positive test instructions, the
negative test instructions, the direct absolute jump instructions, and
the indirect absolute jump instruction are as in \PGLDij.
The effect of a double indirect absolute jump instruction $\diajmp{i}$
is that execution proceeds with the $l$-th instruction of the program
concerned, where $l$ is the content of register $i'$, where $i'$
is the content of register $i$.
If $\diajmp{i}$ is itself the $l$-th instruction, then deadlock occurs.
If $l$ equals $0$ or $l$ is greater than the length of the program,
termination occurs.

Like before, we define the meaning of \PGLDdij\ programs by means of a
function $\pglddijpgldij$ from the set of all \PGLDdij\ programs to the
set of all \PGLDij\ programs.
This function is defined by
\pagebreak[2]
\begin{ldispl}
\pglddijpgldij(u_1 \conc \ldots \conc u_k) = \\ \quad
\psi(u_1) \conc \ldots \conc \psi(u_k) \conc
\ajmp{0} \conc \ajmp{0} \conc
\smash{\overbrace{\ajmp{0}
                   \conc \ldots \conc
                  \ajmp{0}}^{\max(k+2,\maxn)-(k+2)}}
 \conc {} \\ \quad
\ptst{\rf.\eqr{:}1{:}1} \conc \iajmp{1} \conc \ldots \conc
\ptst{\rf.\eqr{:}1{:}n} \conc \iajmp{n} \conc \ajmp{0}
 \conc {} \\ \qquad
  \vdots  \\ \quad
\ptst{\rf.\eqr{:}\maxr{:}1} \conc \iajmp{1} \conc \ldots \conc
\ptst{\rf.\eqr{:}\maxr{:}n} \conc \iajmp{n} \conc \ajmp{0}\;,
\end{ldispl}%
where $n = \min(\maxr,\maxn)$ and the auxiliary function $\psi$ from the
set of all primitive instructions of \PGLDdij\ to the set of all
primitive instructions of \PGLDij\ is defined as follows:
\begin{ldispl}
\begin{aceqns}
\psi(\ajmp{l})   & = & \ajmp{l} & \mif l \leq k\;, \\
\psi(\ajmp{l})   & = & \ajmp{0} & \mif l   >  k\;, \\
\psi(\iajmp{i})  & = & \iajmp{i}\;, \\
\psi(\diajmp{i}) & = & \ajmp{l_i}\;, \\
\psi(u) & = & u  & \mif u\; \mathrm{is\;not\;a\;jump\;instruction}\;,
\end{aceqns}
\end{ldispl}%
and for each $i \in [1,\maxr]$:
\begin{ldispl}
\begin{aeqns}
l_i & = & \maxn + 1 + (2 \mul \min(\maxr,\maxn) + 1) \mul (i - 1)\;.
\end{aeqns}
\end{ldispl}%
The idea is that each double indirect absolute jump can be replaced by
an indirect absolute jump to the beginning of the instruction sequence
\begin{ldispl}
\begin{aeqns}
\ptst{\rf.\eqr{:}i{:}1} \conc \iajmp{1} \conc \ldots \conc
\ptst{\rf.\eqr{:}i{:}n} \conc \iajmp{n} \conc \ajmp{0}\;,
\end{aeqns}
\end{ldispl}%
where $i$ is the register concerned and $n = \min(\maxr,\maxn)$.
The execution of this instruction sequence leads to the intended jump
after the content of the register concerned has been found by a linear
search.
To enforce termination of the program after execution of its last
instruction if the last instruction is a plain basic instruction, a
positive test instruction or a negative test instruction,
$\ajmp{0} \conc \ajmp{0}$ is appended to
$\psi(u_1) \conc \ldots \conc \psi(u_k)$.
Because the length of the translated program is greater than $k$, care
is taken that there are no direct absolute jumps to instructions with a
position greater than $k$.
To deal properly with indirect absolute jumps to instructions with a
position greater than $k$, the instruction $\ajmp{0}$ is appended to
$\psi(u_1) \conc \ldots \conc \psi(u_k) \conc \ajmp{0} \conc \ajmp{0}$ a
sufficient number of times.

Let $P$ be a \PGLDdij\ program.
Then $\pglddijpgldij(P)$ represents the meaning of $P$ as a \PGLDij\
program.
The intended behaviour of program $P$ is the behaviour of $\pglddijpgldij(P)$.
That is, the \emph{behaviour} of $P$, written $\extr{P}_\sPGLDdij$,
is $\extr{\pglddijpgldij(P)}_\sPGLDij$.

The projection $\pglddijpgldij$ uses indirect absolute jumps to obtain
the effect of a double indirect absolute jump in the same way as the
projection $\pgldijpgld$ uses direct absolute jumps to obtain the effect
of an indirect absolute jump.
Likewise, indirect relative jumps can be used in that way to obtain the
effect of a double indirect relative jump.
Moreover, double indirect jumps can be used in that way to obtain the
effect of a triple indirect jump, and so on.

\section{Stack Services}
\label{sect-stack}

In this section, we give a state-based description of the very simple
family of services that constitute a bounded stack of which the elements
are natural numbers up to some bound.
This stack will be used in Section~\ref{sect-PGLDrj} to describe the
behaviour of programs in a variant of \PGLD\ with returning jump
instructions and return instructions.

It is assumed that a fixed but arbitrary number $\maxs$ has been given,
which is considered the greatest length of the stack.
It is also assumed that a fixed but arbitrary number $\maxn$ has been
given, which is considered the greatest natural number that can be an
element of the stack.

The stack services accept the following methods:
\begin{itemize}
\item
for each $n \in [0,\maxn]$, a \emph{stack push method} $\push{:}n$;
\item
for each $n \in [0,\maxn]$, a \emph{stack top test method} $\topeq{:}n$;
\item
a \emph{stack pop method} $\pop$.
\end{itemize}
We write $\Meth_\st$ for the set
$\set{\push{:}n,\topeq{:}n \where n \in [0,\maxn]} \union \set{\pop}$.
It is assumed that $\Meth_\st \subseteq \Meth$.

The methods of stack services can be explained as follows:
\begin{itemize}
\item
$\push{:}n$\,:
if the length of the stack is less than $\maxs$, then the number $n$ is
put on top of the stack and the reply is $\True$;  otherwise nothing
changes and the reply is $\False$;
\item
$\topeq{:}n$\,:
if the stack is not empty and the number on top of the stack is $n$,
then nothing changes and the reply is $\True$; otherwise nothing changes
and the reply is $\False$;
\item
$\pop$\,:
if the stack is not empty, then the number on top of the stack is
removed from the stack and the reply is $\True$; otherwise nothing
changes and the reply is $\False$.
\end{itemize}

Let $s \in \seqof{[0,\maxn]}$ be such that $\len(s) \leq \maxs$.
Then we write $\St_s$ for the service with initial state $s$ described
by
$S =
 \set{\sigma \in \seqof{[0,\maxn]} \where \len(\sigma) \leq \maxs} \union
 \set{\undef}$,
where
$\undef \not\in
 \set{\sigma \in \seqof{[0,\maxn]} \where \len(\sigma) \leq \maxs}$,
and the functions $\eff$ and $\yld$ defined as follows
($n,n' \in [0,\maxn]$, $\sigma \in \seqof{[0,\maxn]}$):%
\footnote
{We write $\seqof{D}$ for the set of all finite sequences with elements
 from set $D$.
 We use the following notation for finite sequences:
 $\emptyseq$ for the empty sequence,
 $\seq{d}$ for the sequence having $d$ as sole element,
 $\sigma \concat \sigma'$ for the concatenation of finite sequences
 $\sigma$ and $\sigma'$, and
 $\len(\sigma)$ for the length of finite sequence $\sigma$.}%
\begin{ldispl}
\begin{gceqns}
\eff(\push{:}n,\sigma)  = \seq{n} \concat \sigma
                                 & \mif \len(\sigma) < \maxs\;,
\\
\eff(\push{:}n,\sigma)  = \sigma & \mif \len(\sigma) \geq \maxs\;,
\\
\eff(\topeq{:}n,\sigma) = \sigma\;,
\\
\eff(\pop,\seq{n} \concat \sigma) = \sigma\;,
\\
\eff(\pop,\emptyseq)              = \emptyseq\;,
\\
\eff(m,\sigma) = \undef          & \mif m \not\in \Meth_\st\;,
\\
\eff(m,\undef) = \undef\;,
\end{gceqns}
\end{ldispl}
\begin{ldispl}
\begin{gceqns}
\yld(\push{:}n,\sigma) = \True   & \mif \len(\sigma) < \maxs\;,
\\
\yld(\push{:}n,\sigma) = \False  & \mif \len(\sigma) \geq \maxs\;,
\\
\yld(\topeq{:}n,\seq{n'} \concat \sigma) = \True  & \mif n = n'\;,
\\
\yld(\topeq{:}n,\seq{n'} \concat \sigma) = \False & \mif n \neq n'\;,
\\
\yld(\topeq{:}n,\emptyseq)               = \False\;,
\\
\yld(\pop,\seq{n} \concat \sigma) = \True\;,
\\
\yld(\pop,\emptyseq)              = \False\;,
\\
\yld(m,\sigma) = \Blocked        & \mif m \not\in \Meth_\st\;,
\\
\yld(m,\undef) = \Blocked\;.
\end{gceqns}
\end{ldispl}%
We write $\St_\mathrm{init}$ for $\St_\emptyseq$.

\section{\PGLD\ with Returning Jumps and Returns}
\label{sect-PGLDrj}

In this section, we introduce a variant of \PGLD\ with returning jump
instructions and return instructions.
This variant is called \PGLDrj.

In \PGLDrj, like in \PGLDij, it is assumed that there is a fixed but
arbitrary finite set of \emph{foci} $\Foci$ with $\st \in \Foci$ and a
fixed but arbitrary finite set of \emph{methods} $\Meth$.
Moreover, we adopt the assumptions made about stack services in
Section~\ref{sect-stack}.
The set $\set{f.m \where f \in \Foci \diff \set{\st}, m \in \Meth}$ is
taken as the set $\BInstr$ of basic instructions.

\PGLDrj\ has the following primitive instructions:
\begin{iteml}
\item
for each $a \in \BInstr$, a \emph{plain basic instruction} $a$;
\item
for each $a \in \BInstr$, a \emph{positive test instruction} $\ptst{a}$;
\item
for each $a \in \BInstr$, a \emph{negative test instruction} $\ntst{a}$;
\item
for each $l \in \Nat$, an \emph{absolute jump instruction} $\ajmp{l}$;
\item
for each $l \in \Nat$,
a \emph{returning absolute jump instruction} $\arjmp{l}$;
\item
an \emph{absolute return instruction} $\return$.
\end{iteml}
\PGLDrj\ programs have the form $u_1 \conc \ldots \conc u_k$, where
$u_1,\ldots,u_k$ are primitive instructions of \PGLDrj.

The plain basic instructions, the positive test instructions, the
negative test instructions, and the absolute jump instructions are as in
\PGLD.
The effect of a returning absolute jump instruction $\arjmp{l}$ is that
execution proceeds with the $l$-th instruction of the program concerned,
but execution returns to the next primitive instruction on encountering
a return instruction.
If $\arjmp{l}$ is itself the $l$-th instruction, then deadlock occurs.
If $l$ equals $0$ or $l$ is greater than the length of the program,
termination occurs.
The effect of a return instruction $\return$ is that execution proceeds
with the instruction immediately following the last executed returning
absolute jump instruction to which a return has not yet taken place.

Like before, we define the meaning of \PGLDrj\ programs by means of a
function $\pgldrjpgld$ from the set of all \PGLDrj\ programs to the set
of all \PGLD\ programs.
This function is defined by
\begin{ldispl}
\pgldrjpgld(u_1 \conc \ldots \conc u_k) = \\ \quad
\psi_1(u_1) \conc \ldots \conc \psi_k(u_k) \conc \ajmp{0} \conc \ajmp{0}
 \conc {} \\ \quad
\ptst{\st.\push{:}1} \conc \ajmp{1} \conc \ajmp{l''}
 \conc \ldots \conc
\ptst{\st.\push{:}1} \conc \ajmp{k} \conc \ajmp{l''} \conc {} \\
\qquad \vdots \\ \quad
\ptst{\st.\push{:}n} \conc \ajmp{1} \conc \ajmp{l''}
\conc \ldots \conc
\ptst{\st.\push{:}n} \conc \ajmp{k} \conc \ajmp{l''}
 \conc {} \\ \quad
\ntst{\st.\topeq{:}1} \conc \ajmp{l''_1} \conc \st.\pop \conc \ajmp{1}
 \conc {} \\
\qquad \vdots \\ \quad
\ntst{\st.\topeq{:}n} \conc \ajmp{l''_n} \conc \st.\pop \conc \ajmp{n}
 \conc {} \\ \quad
\ajmp{l''}\;,
\end{ldispl}%
where $n = \min(k,\maxn)$ and the auxiliary functions $\psi_j$ from the
set of all primitive instructions of \PGLDrj\ to the set of all
primitive instructions of \PGLD\ is defined as follows
($1 \leq j \leq k$):
\begin{ldispl}
\begin{aceqns}
\psi_j(\ajmp{l})  & = & \ajmp{l} & \mif l \leq k\;, \\
\psi_j(\ajmp{l})  & = & \ajmp{0} & \mif l   >  k\;, \\
\psi_j(\arjmp{l}) & = & \ajmp{l_{j,l}}\;, \\
\psi_j(\return)   & = & \ajmp{l'}\;, \\
\psi_j(u) & = & u & \mif u\; \mathrm{is\;not\;a\;jump\;instruction}\;,
\end{aceqns}
\end{ldispl}%
and for each $j \in [1,k]$, $l \in \Nat$, and $h \in [1,\min(k,\maxn)]$:
\begin{ldispl}
\begin{aceqns}
l_{j,l} & = & k + 3 + 3 \mul k \mul ((j - 1) + (l - 1))
 & \mif l \leq k \And j \leq \maxn\;, \\
l_{j,l} & = & j & \mif l \leq k \And j > \maxn\;, \\
l_{j,l} & = & 0 & \mif l > k\;,
\eqnsep
l'      & = & k + 3 + 3 \mul k \mul \min(k,\maxn)\;,
\eqnsep
l''     & = & l' + 4 \mul \min(k,\maxn)\;,
\eqnsep
l''_h   & = & l' + 4 \mul h\;.
\end{aceqns}
\end{ldispl}%
The first idea is that each returning absolute jump can be replaced by
an absolute jump to the beginning of the instruction sequence
\begin{ldispl}
\begin{aeqns}
\ptst{\st.\push{:}j} \conc \ajmp{l} \conc \ajmp{l''}\;,
\end{aeqns}
\end{ldispl}%
where $j$ is the position of the returning absolute jump instruction
concerned and $l$ is the position of the instruction to jump to.
The execution of this instruction sequence leads to the intended jump
after the return position has been put on the stack.
In the case of stack overflow, deadlock occurs.
The second idea is that each return can be replaced by an absolute jump
to the beginning of the instruction sequence
\begin{ldispl}
\begin{aeqns}
\ntst{\st.\topeq{:}1} \conc \ajmp{l''_1} \conc \st.\pop \conc \ajmp{1}
 \conc {} \\
\quad \vdots \\
\ntst{\st.\topeq{:}n} \conc \ajmp{l''_n} \conc \st.\pop \conc \ajmp{n}
 \conc {} \\
\ajmp{l''}\;,
\end{aeqns}
\end{ldispl}%
where $n = \min(k,\maxn)$.
The execution of this instruction sequence leads to the intended jump
after the position on the top of the stack has been found by a linear
search and has been removed from the stack.
In the case of an empty stack, deadlock occurs.
To enforce termination of the program after execution of its last
instruction if the last instruction is a plain basic instruction, a
positive test instruction or a negative test instruction,
$\ajmp{0} \conc \ajmp{0}$ is appended to
$\psi_1(u_1) \conc \ldots \conc \psi_k(u_k)$.
Because the length of the translated program is greater than $k$, care
is taken that there are no non-returning or returning absolute jumps to
instructions with a position greater than $k$.

Let $P$ be a \PGLDrj\ program.
Then $\pgldrjpgld(P)$ represents the meaning of $P$ as a \PGLD\ program.
The intended behaviour of $P$ is the behaviour of $\pgldrjpgld(P)$ on
interaction with a stack.
That is, the \emph{behaviour} of $P$, written $\extr{P}_\sPGLDrj$, is
$\use{\extr{\pgldrjpgld(P)}_\sPGLD}{\st}{\St_\mathrm{init}}$.

According to the definition of the behaviour of \PGLDrj\ programs given
above, the execution of a returning jump instruction leads to deadlock
in the case where its position cannot be pushed on the stack and the
execution of a return instruction leads to deadlock in the case where
there is no position to be popped from the stack.
In the latter case, the return instruction is wrongly used.
In the former case, however, the returning jump instruction is not
wrongly used, but the finiteness of the stack comes into play.
This shows that the definition of the behaviour of \PGLDrj\ programs
given here takes into account the finiteness of the execution
environment of programs.

\section{Conclusions}
\label{sect-concl}

We have studied sequential programs that are instruction sequences with
direct and indirect jump instructions.
We have considered several kinds of indirect jumps, including return
instructions.
For each kind, we have defined the meaning of programs with indirect
jump instructions of that kind by means of a translation into programs
without indirect jump instructions.
Each translation determines, together with some memory device
(a register file or a stack), the behaviour of the programs concerned
under execution.

The increase in the length of a program as a result of translation can
be reduced by taking into account which indirect jump instructions
actually occur in the program.
The increase in the number of steps needed by a program as a result of
translation can be reduced by replacing linear searching by binary
searching or another more efficient kind of searching.
One option for future work is to look for bounds on the increase in
length and the increase in number of steps.

In~\cite{BM06b}, we have modelled and analysed micro-architectures with
pipe\-lined instruction processing in the setting of program algebra,
basic thread algebra, and Maurer computers~\cite{Mau66a,Mau06a}.
In that work, which we consider a preparatory step in the development of
a formal approach to design new micro-architectures, indirect jump
instructions were not taken into account.
Another option for future work is to look at the effect of indirect jump
instructions on pipelined instruction processing.

\bibliographystyle{plain}
\bibliography{TA}

\end{document}